\documentclass[10pt,journal,compsoc]{IEEEtran}
\ifCLASSOPTIONcompsoc
  \usepackage[nocompress]{cite}
\else
  \usepackage{cite}
\fi
\ifCLASSINFOpdf
  \usepackage[pdftex]{graphicx}
\else
\fi
\usepackage{url}

\usepackage{booktabs}
\usepackage{algorithmic}

\usepackage{algorithm2e}

\usepackage{mdframed}
\usepackage{fancybox}
\usepackage{xcolor}
\usepackage{amssymb}

\usepackage{array}
\usepackage{amsmath}

\usepackage{graphicx}
\newtheorem{theorem}{\bf Theorem}

\newenvironment{proof}{\emph{Proof:}}{\hfill$\square$}

\begin{document}
\title{Blizzard: a Distributed Consensus Protocol for Mobile Devices}
%
%
%
%

\author{Mehrdad Kiamari,
        Bhaskar Krishnamachari,
        Muhammad Naveed,
        and~Seokgu Yun
}

\IEEEtitleabstractindextext{%
\begin{abstract}
We present Blizzard, a Byzantine Fault Tolerant (BFT) distributed ledger protocol that is aimed at making mobile devices first-class citizens in the consensus process. Blizzard introduces a novel two-tier architecture by having the mobile nodes communicate through online brokers, and includes a decentralized matching scheme to ensure each node connects to a certain number of random brokers. Through mathematical analysis, we derive a guaranteed safety region (i.e. the set of ratios of malicious nodes and malicious brokers for which the safety is assured) for the Blizzard protocol. Liveness is shown as well. We analyze the performance of Blizzard in terms of its throughput, latency and message complexity. Through experiments based on a software implementation, we show that Blizzard is capable of throughput on the order of several thousand transactions per  second per shard, and sub-second confirmation latency.
\end{abstract}

}

\maketitle

\IEEEdisplaynontitleabstractindextext

%
\IEEEpeerreviewmaketitle

\IEEEraisesectionheading{\section{Introduction}\label{sec:introduction}}

%
%
%
%
\IEEEPARstart{I}{t} has been estimated that there are more than 4 billion mobile device users worldwide~\cite{walden2019introduction} and the mobile ecosystem is still growing~\cite{greenstein2018delivering}. We are therefore motivated to examine the design of a mobile-first consensus protocol suited for distributed ledger maintenance that can leverage such a massive user-base. In particular, in contrast to prior work, that has studied the security and privacy issues around how mobile devices can be used to host client software (i.e. digital wallets) that only send and receive transactions~\cite{biryukov2019security, sai2019privacy}, we aim to examine whether and how these devices can be more directly involved in the consensus mechanism by playing a role in validating transactions. While a major reason for prior work limiting the role of mobile devices to that of client is that they are generally more resource-constrained than larger compute servers, it should be kept in mind that mobile devices today are already capable of a non-trivial amount of  computation, communication and storage and as Moore's law has been shown to apply to mobile platforms as well~\cite{yu2019moore}, these capabilities will only continue to grow into the future. 

The distributed mobile-based consensus protocol for distributed ledger maintenance that we propose is called Blizzard. Blizzard is a leaderless consensus protocol in which mobile nodes each connect and communicate with a number of online servers called brokers. Mobile nodes only need to store and communicate with a number of end addresses that scales with the number of brokers in the system, rather than with the number of all mobile devices in the system\footnote{Of course, non-mobile devices such as online servers could also participate as validators, our point is that this is the first protocol to explicitly allow for mobile device-based validators, which can be switched off or connect intermittently.}.  Each mobile node connects to a random subset of these servers for  a given period of time and communicates with all other mobile nodes in each broker's group.  Effectively, each broker creates a broadcast group of mobile nodes that can query and respond to each other.

We briefly enumerate our contributions in this work as follows:
\begin{enumerate}
    \item Blizzard is the first mobile-based leaderless Byzantine Fault Tolerant (BFT) distributed consensus protocol. Not only can mobile devices issue transactions, but they can participate in the core transaction verification and consensus  process as well. This could increase by 2-3 orders of magnitude the number of nodes that can participate in validation, improving both adoption and security.
    \item We propose a novel two-tier protocol, where consensus between mobile nodes is enabled by the use of online brokers, and a decentralized matching scheme that ensures each mobile node connects to $k$ random brokers. 
    \item Provable safety guarantee. 
    We mathematically derive the set of  ratios of Byzantine nodes and brokers for which Blizzard's safety is guaranteed. We also discuss why liveness holds in Blizzard. 
    

    \item We analytically characterize the throughput of Blizzard by modeling the sequential pipeline involved in processing transactions at each node and identifying the throughput bottleneck via empirical profiling. 
    
    \item Likewise, we also analytically characterize the confirmation latency and message complexity of Blizzard. We show that the use of brokers in Blizzard creates a communication topology that allows for efficient consensus; under reasonable parameter settings, transactions are propagated in  Blizzard within just 4 communication rounds with high probability. 
    
    \item Through experiments based on a software implementation of Blizzard, we show that it is capable of 10,000 transactions per second per shard, and allows for sub-second confirmations. 
    
\end{enumerate}
    

\section{Related Works}

\begin{table*}
\caption{Comparison of Different Protocols.}
\begin{tabular}{ |p{1.5cm}||p{2.5cm}|p{1cm}|p{1.5cm}|p{.75cm}| p{2cm}| p{2.5cm}|p{1.5cm}|p{.75cm}|}
 \hline
 Protocol &Sybil Control Method& Leaderless &Ledger Structure & BFT & Transaction per Second per shard & Confirmation Latency& Number of Validators & Mobile-based\\
 \hline
 \hline
 Bitcoin   & PoW    &No&   Chain & No & 7&$\sim$ 40mins& 100k+ &No \\
 \hline
 Ethereum&   PoW/PoS  & No   &Chain&No & 20& $\sim$ 60s & 100k+ & No\\
 \hline
 Tendermint &Agnostic & No&  Chain&Yes & $\sim$ 10,000& $\sim$ 2-15s& $\sim$100-1k & No\\
 \hline
 Avalanche  &Agnostic & Yes&  DAG&Yes & $\sim$ 3,400& $\sim$ 1.35s&100k+ & No \\
 \hline
 Blizzard   &Agnostic & Yes&  DAG&Yes & $\sim$ 10,000& $\sim$ 0.65s &{\bf 100M+} & Yes\\
 \hline
\end{tabular}
\label{table:diff_protocols}
\end{table*}

The original Bitcoin paper by Nakomoto~\cite{Bitcoin} provided a joint solution to several problems including consensus (through longest-chain adoption), ledger representation (hashed chain of blocks), Sybil control (through Proof of Work), and incentives (through mining and transaction rewards). However, it is possible, and we argue, of value, to consider these various components separately. We briefly survey the literature with respect to these four  dimensions to place our work in context.  

{\noindent{{\bf Consensus}}:} In this work, we focus solely on the problem of distributed consensus, focusing in particular on BFT. In~\cite{Backbone}, the Bitcoin protocol is formalized and it is shown how it can be extended to provide Byzantine agreement. There is in fact an earlier literature on BFT consensus for distributed systems, primarily focused on leader-based protocols, such as PBFT~\cite{PBFT} and BFT-Smart~\cite{BFTsmart} and others. More recently, in the context of Blockchain, several new leader-based consensus solutions have been proposed that improve the speed of BFT consensus under partial synchrony assumptions -- these include Tendermint~\cite{Tendermint}, Hotstuff~\cite{Hotstuff} and Casper CBC~\cite{CasperCBC}. There has also been recent work on leaderless protocols, including Hashgraph~\cite{Hashgraph}, Avalanche~\cite{avalanche19}, DBFT~\cite{DBFT} and Aleph~\cite{Aleph}, which do not require that there be a single proposer or leader for each round of consensus. The Blizzard protocol described in this paper is a leaderless BFT protocol that is designed to allow mobile devices to participate in the consensus by leveraging online brokers. Blizzard builds on the idea of gossip-based consensus presented originally in Avalanche but is structurally significantly different due to the introduction of aggregating Brokers and therefore requires a different safety, throughput and latency analysis which are all presented in this work.

{\noindent{{\bf Ledger representation}}:} There are broadly two classes of approaches to ledger representation in distributed ledger systems - either using a linear Blockchain as in the original Bitcoin, Ethereum and many other protocols, or allowing transactions (or blocks) to point to other previous transactions (or blocks) resulting in a directed acyclic graph data (DAG) structure. Examples of such DAG-based protocols include IOTA~\cite{Iota}, Hashgraph~\cite{Hashgraph}, Avalanche~\cite{avalanche19}, and Helix~\cite{Helix}. Blizzard follows the same DAG-based approach proposed in Avalanche. 

{\noindent{{\bf Sybil control}}:} Surveying different Sybil control mechanisms, we find that Ethereum also uses Proof of Work for Sybil control while newer projects and systems such as Cosmos~\cite{Cosmos}, Algorand~\cite{Algorand}, Ouroboros~\cite{Ouroboros}, Dfinity~\cite{Dfinity} and Ethereum 2.0~\cite{Eth2} have been exploring energy-efficient alternatives such as Proof of Stake and delegated Proof of Stake. Meanwhile in permissioned blockchains such as Hyperledger Fabric~\cite{Fabric} and Hyperledger Sawtooth~\cite{Sawtooth}, Sybil control is handled explicitly by only allowing a limited set of pre-vetted, pre-approved validators into the system. In this work, we do not treat the problem of Sybil control, similar to other prior work focused on distributed consensus. 

{\noindent{{\bf Incentives}}:} Finally, like most prior work on BFT consensus protocols and permissioned blockchains (but unlike many cryptocurrency projects such as Bitcoin, Ethereum), Blizzard is agnostic to how incentives are provided to validating nodes - this allows its use for a broader range of use cases beyond cryptocurrency, but keeps the flexibility to allow incentive mechanisms to be employed as a separate modular layer if needed. 

One crucial aspect of widely adopted protocols such as Bitcoin, Ethereum and Avalanche is that they are open, permissionless protocols, meaning that any device can join the network or leave them at any time. Therefore they do  not utilize any information about the total number of validator nodes when voting for blocks. This significant feature, which is also essential for mobile networks, incurs the cost that such a validator number-agnostic protocol can only be proved to operate safely under a synchronous model~\cite{permissionless_sync}. In particular, \cite{permissionless_sync} has shown that permissionless consensus cannot be proved safe in partially synchronous or asynchronous models. The Blizzard protocol we present here also does not make assumptions about the \emph{total} number of participating nodes and hence is also restricted to a synchronous model with respect to its safety guarantee  (as is the case with Bitcoin, Ethereum and Avalanche).    

In Table \ref{table:diff_protocols}, we summarize some of the main Blockchain protocols and their key properties and attributes along with Blizzard, to help put our contribution in context. In a nutshell, as we will show, Blizzard provides both high throughput and low latency comparable to state of the art protocols, while allowing significantly greater scale because it is designed to allow mobile devices to serve as transaction validators.

\section{System Model}
\begin{figure*}[t]
\centering
\includegraphics[trim=70mm 63mm 70mm 45mm,clip,width=.7\textwidth]{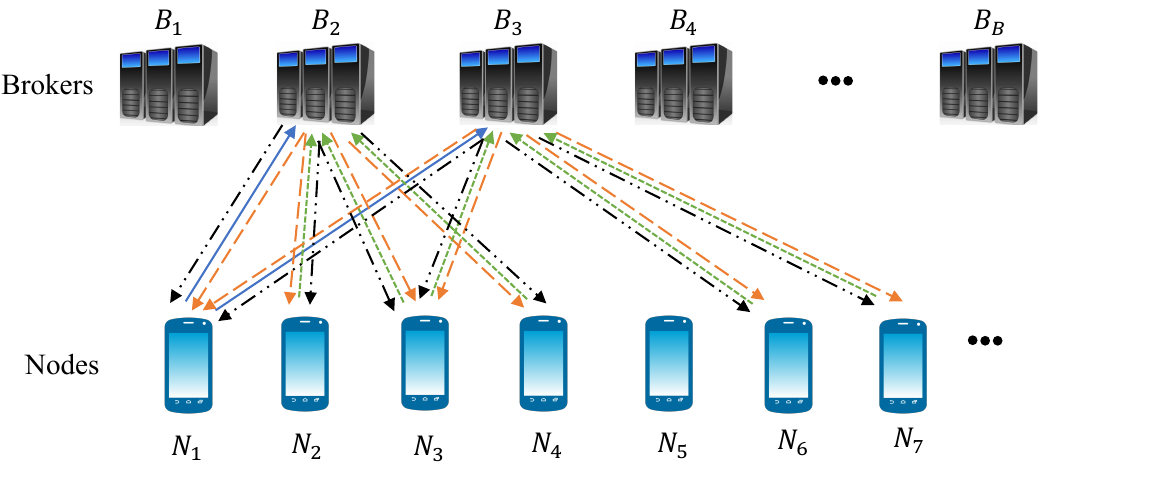}
\caption{An illustration on how Blizzard works for $k=2$. Mobile node 1 queries brokers $B_2$ and $B_3$ on a new transaction these brokers have not queried yet (shown with blue arrows). Then, these brokers queries all their connected mobile nodes about the transaction (depicted with orange arrows). Afterwards, all connected mobile nodes respond back to queries (shown with green arrows) and brokers reflect the majority vote to all of the connected nodes (presented by dashed-black arrows).}
\label{fig_model}
\end{figure*}

We consider a broker-assisted mobile network where disjoint sets ${\mathcal N}_C$ and ${\mathcal N}_M$ respectively represent sets of correct and malicious mobile nodes (set of all nodes is denoted by $\mathcal N := {\mathcal N}_C \cup {\mathcal N}_M$). Furthermore, set $\mathcal B$ indicates set of all brokers, which consists of sets ${\mathcal B}_C$ and ${\mathcal B}_M$ representing the subsets of correct and malicious brokers, respectively. 
Other notations are provided in Table \ref{tab:TableOfNotation}.
\begin{table}[htbp]\caption{Notations Description}
\begin{center}
\begin{tabular}{r c p{7cm} }
\toprule
$n$ &:& Number of all mobile nodes (where $n=|\mathcal N|$).\\
$c$ &:& Number of correct mobile nodes.\\
$b$ &:& Number of Byzantine mobile nodes.\\
$m$ &:& Number of all brokers.\\
$m_c$ &:& Number of correct brokers.\\
$m_b$ &:& Number of Byzantine brokers.\\
${\mathcal N}^{(b)}$ &:& Set of connected mobile nodes to broker $b$.\\
${\mathcal B}^{(u)}$ &:& Set of connected brokers to mobile node $u$.\\
$k$ &:& Number of brokers being sampled by each mobile node.  \\
$\alpha$&: & Majority threshold of mobile nodes for considering a ``yes'' vote.  \\
$\eta$&: & Majority threshold of brokers for considering a ``yes'' vote.  \\
$\beta_1$ &:& Security threshold used for consecutive counter.\\
$\beta_2$ &:& Security threshold used for confidence counter.\\
\bottomrule
\end{tabular}
\end{center}
\label{tab:TableOfNotation}
\end{table}

Mobile nodes issue cryptographically signed transactions. We assume all validating nodes have access to a common function that can determine if any two transactions are conflicting or not\footnote{This is general enough, for example, to cover the detection of conflicting transactions under either a UTXO or account-based model.}. Correct nodes never issue conflicting transaction, while Byzantine nodes may issue conflicting transactions. 

Regarding misbehavior acts, we assume the existence of both Byzantine mobile nodes as well as brokers. In Blizzard, malicious brokers are effectively limited to suppressing messages as they do not sign or initiate any messages themselves and are assumed to not being able to forge messages from mobile nodes. 
As far as the Byzantine behavior for malicious mobile nodes is concerned, malicious nodes are computationally limited (not able to forge signature) while they can choose any execution strategy that they desire. Moreover, mobile nodes being/switching off does not affect the consensus as long as the fraction of correct connected mobile nodes are sufficiently high. 

Since we aim to have a protocol such that any vote on a new transaction would be a vote on some previous transactions, we incorporate a DAG structure into our protocol. To do so, some parent transactions would be assigned for each new transaction. Therefore, any vote on a specific transaction is a vote on all of its ancestor transactions\footnote{All transactions accessible through the parent of a transaction are referred as \emph{ancestor transactions}.} as well. 
The overview of the DAG structure of Avalanche is presented in Appendix B.

\small
\begin{algorithm}\label{main_alg}
\caption{Blizzard Algorithm}
\SetAlgoLined
{\bf Initialization:}\\
- Each node $u\in {\mathcal N}$ randomly connects to $k$ \\brokers represented by ${\mathcal B}^{(u)}$.\\ 
- Set ${\mathcal T}_u={\mathcal Q}_u = \emptyset$ for all nodes which do not issue transaction where ${\mathcal T}_u$ and ${\mathcal Q}_u$ represent known and queried transaction sets of node $u$, respectively.\\
 \While{there is a transaction $T$ at any node $u$ such that $T\in{\mathcal T}_u, T\not\in{\mathcal Q}_u$}{
  - $R_{\text{brokers}}:= \sum_{b\in {\mathcal B }^{(u)}}{Query_{Broker}(b,T)}$\;
  \If{$R_{\text{brokers}}\geq \alpha k$}{
  - $v_{u,T}=1$\ // $T$ receives \emph{voucher} and \\\hspace{20mm}appended to DAG of node $u$.
  \\- Update DAG and conflicting sets of node $u$ after appending $T$.
  }{
  - ${\mathcal Q}_u = {\mathcal Q}_u \cup \{T\}$  // mark $T$ as queried transaction.}}
\end{algorithm}

\begin{equation}\label{side_funcs}
Query_{_{Broker}}(b,T):=
\left\{
	\begin{array}{ll}
		1  &  \mbox{if } \sum_{u'\in {\mathcal N}^{(b)}}{Query_{Node}(u',T)} \geq \\ &  
		~~~~~~~~~~~~~~~~~~~~\eta |{\mathcal N}^{(b)}| \\
		0 & \mbox{if } \text{else}
	\end{array}
\right.
\end{equation}

\normalsize   

We next present how our proposed Blizzard scheme works in detail. 

\subsection{Proposed Blizzard Scheme}
Our proposed Blizzard scheme works as follows:
each node $u\in \mathcal N$ connects to $k$ brokers (represented by $\mathcal B^{(u)}$) uniformly at random (the mechanism of $k$ random connections will be discussed in the following subsection) and queries them on a new transaction $T$. 

Upon receiving a query, each broker $b\in  {\mathcal B}^{(u)}$ computes $\eta$-majority vote on $T$ by querying all of its connected nodes denoted by ${\mathcal N}^{(b)}$. Then, each node $v\in {\mathcal N}^{(b)}$ for all broker $b\in{\mathcal B}^{(u)}$ affirmatively responds to the query if all of the ancestor transactions of $T$ are currently the preferred choice of transaction in their corresponding conflict sets in the stored DAG of node $v$. Afterwards, broker $b \in  {\mathcal B}^{(u)}$ aggregates the count of all positive responses collected from nodes ${\mathcal N}^{(b)}$ and sends an affirmative response to all its connected nodes using a suitable key aggregation scheme if $\geq \eta |{\mathcal N}^{(b)}|$ (where $\frac{1}{2}<\eta<1$) collected responses from nodes ${\mathcal N}^{(b)}$ are positive. In the case of having $\geq \alpha k$ (where $\frac{1}{2}<\alpha<1$) positive responses, collected from brokers $\mathcal B_u$ for $T$, then $T$ will be appended to the stored DAG of node $u$ and node $u$ never queries $T$ again. The protocol would continue by following the same process for all nodes having $T$ in their known transaction sets. An example of our proposed scheme for $k=2$ where one node queries on a new transaction is depicted in Fig. \ref{fig_model}. The details of Blizzard protocol are elaborated in Algorithm \ref{main_alg} and side function $Query_{broker}(.,.)$ in  (\ref{side_funcs}). 
Note that only side function $Query_{node}(.,.)$ and functions having to do with finalizing a transaction are the same as Avalanche. 

We next present a distributed mechanism to enforce $k$ random connections for mobile nodes.

\subsection{Distributed Random Matching}
For Blizzard to work, we need to ensure that each mobile node is connected to $k$ \emph{random} brokers. 
Since random connection plays a key role in preventing collusion between Byzantine entities, we propose a mechanism which requires all mobile nodes, even Byzantine ones, to provide a proof of their connections being random.

Our proposed distributed random matching scheme works as follows: 
\begin{enumerate}
    \item Each mobile device applies a Hash function on the combination of the random number coming from a distributed random beacon\footnote{Note that a distributed random beacon is now live online at \url{https://drand.love/} .}~\cite{DRB} with ID of the mobile device. Regarding the Hash function, it outputs $B$ bits where $0$ and $1$ are equally likely. Then the indices of the first $k$ ones represent the brokers each mobile device has to connect with. The mobile device afterwards sends the output of its Hash function as well as its IDs to the brokers it is supposed to connect with.
    \item Brokers validate the ID of mobile devices, verify that hash values generated by mobile devices are correctly produced, taking into account the distributed random beacon, and thus verify that mobile devices are authorized to connect.
\end{enumerate}

\begin{figure}[t]
\centering
\includegraphics[trim=50mm 80mm 110mm 27mm,clip,width=.5\textwidth]{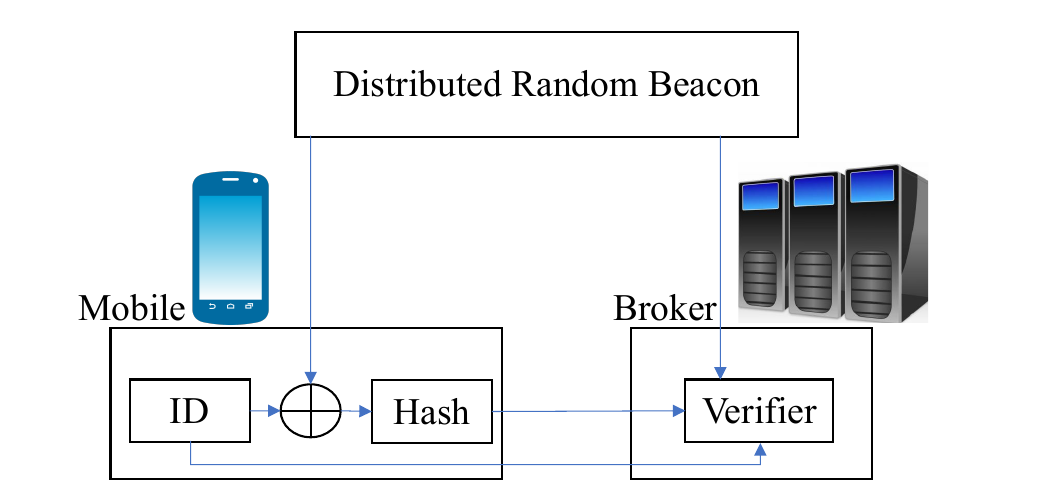}
\caption{Proposed distributed random matching scheme for connecting each mobile node to $k$ brokers.}
\label{DistributedRandomBeacon}
\end{figure}

An illustration of proposed distributed random matching scheme is depicted in Fig. \ref{DistributedRandomBeacon}.
One crucial point guaranteeing our proposed scheme works is that the Hash function outputs sufficiently long sequence (or equivalently large $B$) such that there exists at least $k$ ones in the hashed value with high probability.  
Theorem \ref{Thm_random_matching} determines parameter $B$ for our proposed distributed random matching scheme to satisfy this condition.

\begin{theorem}\label{Thm_random_matching}
In order to have the probability of existing at least $k$ ones in a sequence of length $B$ to be $1-\delta$ (for small $\delta$), parameter $B$ should satisfy $\frac{1}{2}\log \frac{1}{\delta} = (\frac{1}{2}-\frac{k-1}{B})^2B$.
\end{theorem}
\begin{proof}
\small
\begin{equation}\label{Thm_random_matching_eq}
\begin{aligned}
P(H(B)\leq k-1) \leq \delta=\exp{(-2\epsilon^2B)}
\end{aligned}    
\end{equation}
\normalsize
where (\ref{Thm_random_matching_eq}) follows from Hoeffding's inequality,  $H(B)$ indicates the number of ones in a sequence of length $B$, and $\epsilon := \frac{1}{2}-\frac{k-1}{B}$.   
\end{proof}

\section{Safety and Liveness}

As mentioned earlier, it has been shown in~\cite{permissionless_sync} that permissionless protocols that do not make any assumption about the total number of nodes involved in the consensus protocol can only be proved to operate correctly under a synchronous model. This applies to previously proposed protocols such as Bitcoin, Ethereum and Avalanche, and also to our Blizzard protocol. Thus, in the following, we are assuming a network that is synchronous in that the maximum latency for any message in the network is bounded by a known constant.

\subsection{Safety Analysis}

To  show  the  safety  of Blizzard scheme, it  is enough to show that all correct nodes decide on the same transaction among all conflicting transactions in finite time almost  surely. To represent preference of nodes for two conflicting transactions, we let nodes to take two different colors, namely red and blue. 
Nodes which prefer transactions $T_1$ and $T_2$ can be represented with nodes with red and blue colors, respectively. Without  loss  of  generality, we  can  only  focus  on  the  case  where  all correct nodes reach consensus on red color.

The behavior of all mobile nodes and brokers are as follows:
\begin{enumerate}
\item Every correct mobile node always responses with honesty upon receiving any query. 

\item A queried Byzantine node may respond with any color or even refuse to respond. 

\item Every correct broker computes $\eta$-majority of the votes collected from all nodes connected to it and \emph{broadcasts} the majority vote to \emph{all} of them. 

\item Byzantine brokers cannot forge information since they cannot cryptographically sign transactions from nodes. However, a Byzantine broker may compute $\eta$-majority of the votes collected from any \emph{subset} of nodes connected to it and send the computed vote to any selected nodes which are connected to this broker. \end{enumerate}

In order to prevent adversarial attacks, similar to Avalanche, we incorporate two following counters for each mobile node:

1) \emph{Conviction in Current Color (C3)} counter to store how many consecutive computed majority votes have resulted in the same color. Once a node flips its color, this counter reset to zero. Furthermore, a node locks into the current color when this counter exceeds some security parameter $\beta_1$.

2) \emph{Confidence} counter to take into account the number of queries which have yielded a majority vote for their corresponding colors. A node flips its color only if the confidence value of its computed vote is larger than the confidence value of the current color. Moreover, a node locks into a color once the confidence value of this color exceeds some security parameter ${\beta}_2$.

\noindent{\bf Color-based Blizzard:}
Considering the aforementioned assumptions, the color-based Blizzard scheme works as follows: each mobile node connects to $k$ brokers uniformly at random and queries them. 
Please note that randomly connecting each mobile to $k$ brokers is guaranteed by the  distributed random matching scheme described in the previous section.
Then, each of these brokers queries all its connected nodes regarding their colors. Subsequently, each of the connected nodes, upon being queried, sends its color to the brokers which it is connected to. Once the color of all nodes connected to broker $i$ received, then the broker computes whether $\geq \eta |{\mathcal N}^{(i)}|$ (where ${\mathcal N}^{(i)}$ represents nodes connected to broker $i$ and $\eta \in (\frac{1}{2},1]$) collected responses have the same color or not. In the case of having $\geq \eta |{\mathcal N}^{(i)}|$ responses with the same color, then the broker broadcasts the computed majority vote to all nodes connected to it. 

Once $\alpha$-majority of votes collected from brokers connected to a node yields to a color, i.e. receiving $\geq \alpha k$ positive responses, the Confidence counter for that color will increase by one. If this color (the one computed from the majority votes coming from $k$ brokers) is the same as the node's current color, C3 counter increases by one; otherwise the node resets C3 counter to zero.
A node flips its color to a new color if the Confidence counter of new color is larger than the Confidence counter of current color. 

\noindent{{\bf Safety Analysis of Color-based Blizzard:}}
Without loss of generality we assume that the initial node colors are randomly assigned such that $\frac{c}{2}+1$ nodes have red color while the remaining nodes are blue \footnote{This is the worst case scenario for reaching consensus due to balance in number of nodes with different colors.}. 

Let us represent correct mobile nodes which prefer red and blue colors with $u$ and $v$, respectively.
To show consensus is achieved, we can show that nodes preferred blue color acquire confidence in red color as time elapsed and they eventually turn into red-preferred nodes with high probability. Note that node $z$ change its color to red if $z.\text{Conf}[R]>z.\text{Conf}[B]$ and flips to blue otherwise.  

Our proof for reaching consensus will be shown as the following steps: 

{\noindent {\bf Step 1}:} After some finite time, system reaches to the point where there are $\frac{c}{2}+\Delta$ red nodes while the remaining nodes are blue.

{\noindent {\bf Step 2}:} At this point, $v$ nodes have negative average growth in confidence value for blue color at any time \footnote{Average growth in confidence value of node $z$ for color $x$ is ${\mathbb E}[z.\text{Conf}^{~(t)}[x]]-{\mathbb E}[z.\text{Conf}^{~(t-1)}[x]]$.}, with high probability. After short period of time, we have $v.\text{Conf}[B] - v.\text{Conf}[R]=-1$, with high probability, and as a result, $v$ nodes flips their color to red. Note that $u$ nodes just gain more confidence in red as time elapses in this step.   

\subsubsection{Step 1}
We can model our scheme as a discrete time Markov Chain with state $s_i$, $\forall i\in\{0,\ldots,c\}$, where $s_i$ represents the state with $i$ red and $c-i$ blue nodes, with transition probability matrix $M$. 

Since each of these transition probabilities are a function of the adversary, let us now elaborate upon its  behavior. The most malicious scenario conducted by adversary aims to achieve the following goal: keep the confidence value of blue and red colors nearly the same for $u$ nodes, while letting $v$ nodes to increase their confidence value of blue color as much as they can. More formally, the scenario is to have $u.\text{Conf}[R] = u.\text{Conf}[B]+1$ and maximizing $\kappa$ where $\kappa\triangleq  v.\text{Conf}[B] - v.\text{Conf}[R]$. Therefore,

\noindent{$\bullet$ \bf When a $u$ node queries}:
All Byzantine mobile nodes acquire red colors. Regarding the malicious brokers, all of them act with honesty without suppressing any color. 

\noindent{$\bullet$ \bf When a $v$ node queries}:
All Byzantine mobile nodes pick blue colors and all Byzantine brokers with red-color majority switch to a blue-majority broker by not reflecting their red-color mobile nodes. With high probability, as shown in Appendix A, a node is connected to at most $f\triangleq \frac{m_br}{m}$ Byzantine brokers in a population of $r$ red brokers and $m-r$ blue brokers.  

Since none of the transition probabilities are zero, the system can reach to the state $s_{c/2+\Delta}$ in finite time. We will discuss how $\Delta$ can be determined in the next step. 
It is important to emphasize that we set the security parameters $k$, $\alpha$,$\eta$, $\beta_1$, and $\beta_2$ such that no node finalizes its color during this step.     

\subsubsection{Step 2}
In this step, we show that how $v$ nodes flip their color to red and as a result, all correct nodes reach consensus with high probability. To do so, we write the expected confidence value for mobile nodes at time $t$ as follows:
\small
\begin{equation}\label{ur}
\begin{aligned}
&{\mathbb E}[u.\text{Conf}^{~(t)}[R]]= {\mathbb E}[u.\text{Conf}^{~(t-1)}[R]]+ P({\mathcal C}_u^{(t)} \text{is red})
\\&={\mathbb E}[u.\text{Conf}^{~(t-1)}[R]]+ \sum_{r=\alpha k}^m {P({\mathcal C}_u^{(t)}=\text{red}~|~{{\mathcal A}_{r}^i})
P({\mathcal A}_{r}^i)}
\\&={\mathbb E}[u.\text{Conf}^{~(t-1)}[R]]  
\\&~~~~~+\sum_{r=\alpha k}^{m}
\Big(\sum_{j'=\alpha k}^{k}{\frac{ {{r}\choose{j'}} {{m-r}\choose{k-j'}}   }{{{m}\choose {k}}}}\Big)
{{m}\choose{r}}p_{i+b}^r {(1-p_{i+b})}^{m-r},
\end{aligned}    
\end{equation}
\normalsize
where ${\mathcal C}_u^{(t)}$ represents the color of the computed vote of node $u$ at time $t$. Moreover, ${\mathcal A}_{r}^i$ denotes the event of having $r$ red-majority brokers and $m-r$ blue-majority brokers when there are $i$ red mobile nodes in a population of $n$ nodes. Therefore, $P({\mathcal A}_{r}^i) = {{m}\choose{r}}p_i^r {(1-p_i)}^{m-r},$
where $p_i$, the probability of a broker to be red given total $i$ red nodes in a population of $n$ nodes, is
\small
\begin{equation}\label{red_broker_prob}
\begin{aligned}
p_i &\triangleq \sum_{\ell=1}^n{ 
\underbrace{\frac{\sum_{j\geq {\eta \ell}}{{{i}\choose{j}}{{n-i}\choose{\ell-j}}}}{ {{n}\choose{\ell}}  }.}_{\text{probability of the broker being red given}~\ell~\text{  connections}}}
\\&~~~~\times\underbrace{{{n}\choose{\ell}}
{(\frac{1}{m})}^{\ell}{(1-\frac{1}{m})}^{n-\ell}
}_{\text{probability of the broker having}~\ell~\text{connections}}
\end{aligned}
\end{equation}
\normalsize

Similarly, we would have
\small
\begin{equation}\label{ub}
\begin{aligned}
{\mathbb E}&[u.\text{Conf}^{~(t)}[B]]= {\mathbb E}[u.\text{Conf}^{~(t-1)}[B]]
\\&~~~~~+ \sum_{r=\alpha k}^{m}
\Big(\sum_{j'=\alpha k}^{k}{\frac{ {{r}\choose{j'}} {{m-r}\choose{k-j'}}   }{{{m}\choose {k}}}}\Big)
{{m}\choose{r}}{(1-p_{i})}^{r}{p_{i}}^{m-r}
\end{aligned}    
\end{equation}

\begin{equation}\label{vr}
\begin{aligned}
{\mathbb E}&[v.\text{Conf}^{~(t)}[R]]= {\mathbb E}[v.\text{Conf}^{~(t-1)}[R]]
\\&~~~~~+ \sum_{r=\alpha k}^{m}
\Big(\sum_{j'=\alpha k}^{k}{\frac{ {{r-f}\choose{j'}} {{m-r+f}\choose{k-j'}}   }{{{m}\choose {k}}}}\Big)
{{m}\choose{r}}p_{i}^r {(1-p_{i})}^{m-r}
\end{aligned}    
\end{equation}

\begin{equation}\label{vb}
\begin{aligned}
{\mathbb E}&[v.\text{Conf}^{~(t)}[B]]= {\mathbb E}[v.\text{Conf}^{~(t-1)}[B]]
\\&~~~~~+ \sum_{r=\alpha k}^{m}
\Big(\sum_{j'=\alpha k}^{k}{\frac{ {{r+f}\choose{j'}} {{m-r-f}\choose{k-j'}}   }{{{m}\choose {k}}}}\Big)
{{m}\choose{r}}{(1-p_{i})}^{r}{p_{i}}^{m-r}.
\end{aligned}    
\end{equation}
\normalsize

By defining $D\triangleq \Big({\mathbb E}[v.\text{Conf}^{~(t)}[B]]- {\mathbb E}[v.\text{Conf}^{~(t-1)}[B]]\Big) - \Big({\mathbb E}[v.\text{Conf}^{~(t)}[R]]-{\mathbb E}[v.\text{Conf}^{~(t-1)}[R]]\Big)$, we have
\small
\begin{equation}\label{I}
\begin{aligned}
D &= \sum_{r=\alpha k}^{m}
\Big(\sum_{j'=\alpha k}^{k}{\frac{ {{r+f}\choose{j'}} {{m-r-f}\choose{k-j'}}   }{{{m}\choose {k}}}}\Big)
{{m}\choose{r}}{(1-p_{i})}^{r}{p_{i}}^{m-r}
\\&~~-
\sum_{r=\alpha k}^{m}
\Big(\sum_{j'=\alpha k}^{k}{\frac{ {{r-f}\choose{j'}} {{m-r+f}\choose{k-j'}}   }{{{m}\choose {k}}}}\Big)
{{m}\choose{r}}p_{i}^r {(1-p_{i})}^{m-r}.
\end{aligned}    
\end{equation}
\normalsize
We now aim to show that $D$ 
acquires negative values and once $v.\text{Conf}[B]-v.\text{Conf}[R]$ reaches value $-1$, all correct nodes acquire red color. 
Let us introduce the following random variables $X_{t} \triangleq v.\text{Conf}^{~(t)}[B]-v.\text{Conf}^{~(t)}[R]$ and $X_{1:t} \triangleq \sum_{i=1}^t{X_i}$.

Since $X_{1:t}$ satisfies Hoeffding's inequality condition due to the fact that $a)$  $~X_t$ are i.i.d and $~b)$ $~X_t$'s are sub-Gaussian because of taking bounded values, we would have $P(X_{1:t}-{\mathbb E}[X_{1:t}]\geq q) \leq \exp{(-2tq^2)}$.

Therefore, in order to show $X_{1:t}$ is negative, with high probability, it suffices to show that ${\mathbb E}[X_{1:t}]$ is negative. To do so, based on the recursive formula  (\ref{vr}) and (\ref{vb}), we need to show that $D$ acquires negative values under certain conditions. Let us first elaborate upon how (\ref{I}) can be approximated as in the following Theorem. 

\begin{theorem}\nonumber
$D$ can be approximated as follows\end{theorem}
\small
\begin{equation}\label{approximation_I}
\begin{aligned}
D ~{\approx} ~
G(k,m,1+\rho_b,\alpha,1-p_i)-
G(k,m,1-\rho_b,\alpha,p_i),
\end{aligned}    
\end{equation}
\normalsize
where 
\small
\begin{equation}\label{def_apprx}
\begin{aligned}
G(k,m,\lambda,\alpha,p_i)&\triangleq
\sum_{r} \frac{1}{1+e^{-1.702\Big(\frac{\frac{k}{m}\lambda r -\alpha k}{\sqrt{\frac{k}{m}\lambda r (1-\frac{\lambda r}{m})}} \Big)}}
\frac{e^{-\frac{(r-mp_i)^2}{2\sigma_i^2}}}{\sqrt{2\pi \sigma_i^2}},
\\\sigma_i^2 &\triangleq mp_i(1-p_i).
\end{aligned}    
\end{equation}
\normalsize
\begin{proof}
We note that the Probability Mass Function (PMF) of hyper-geometric distribution with parameters $(r,m,k)$ is $p_{hg}(x;r,m,k)\triangleq {\frac{ {{r}\choose{x}} {{m-r}\choose{k-x}}   }{{{m}\choose {k}}}}$. By approximating 
\begin{enumerate}
    \item Binomial distribution $Bionomial(n,p)$ (corresponds to terms ${{m}\choose{r}}p_{i}^r {(1-p_{i})}^{m-r}$ or ${{m}\choose{r}}{(1-p_{i})}^{r}{p_{i}}^{m-r}$) with Normal distribution ${\mathcal N}(np,np(1-p))$,
    \item Hyper-geometric distribution $Hyper-Geometrical(r,m,k)$ with Normal distribution ${\mathcal N}(k\frac{r}{m},k\frac{r}{m}(1-\frac{r}{m}))$,
\end{enumerate}
 $D$ can be approximated as 
 \small
\begin{equation}\label{approximation_I_proof}
\begin{aligned}
D &\approx \sum_{r}
\Big(1-\Phi(\frac{\alpha k - k\frac{r+f}{m}}{\sqrt{k\frac{r+f}{m}(1-\frac{r+f}{m})}}) \big)
\frac{e^{-\frac{(r-m(1-p_i))^2}{2\sigma_i^2}}}{\sqrt{2\pi \sigma_i^2}} 
\\&~~-
\sum_{r}
\Big(1-\Phi(\frac{\alpha k - k\frac{r-f}{m}}{\sqrt{k\frac{r-f}{m}(1-\frac{r-f}{m})}}) \big)
\frac{e^{-\frac{(r-mp_i)^2}{2\sigma_i^2}}}{\sqrt{2\pi \sigma_i^2}}.
\end{aligned}    
\end{equation}
\normalsize
where $\Phi(x)$ represents the Cumulative Distributive Function (CDF) of Normal distribution $\mathcal N(0,1)$. According to \cite{approx_with_logistic}, $\Phi(x) \approx \frac{1}{1+e^{-1.702x}}$. Therefore, by substituting $f=\rho_b r$ and using the aforementioned approximation of $\Phi(x)$, we can approximate $D$ as (\ref{approximation_I}). 
\end{proof}

\noindent{\bf Remark:}
One can easily see that $D$ can acquire negative values when $p_i > \frac{1}{2}$. This is due to the fact that the logistic coefficient of Normal distribution in $G(.)$ considerably scales down Normal distribution ${\mathcal N}(m(1-p_i),mp_i(1-p_i))$ (appeared in $G(k,m,1+\rho_b,\alpha,1-p_i)$) compared to Normal distribution ${\mathcal N}(mp_i,mp_i(1-p_i))$ (appeared in $G(k,m,1-\rho_b,\alpha,p_i)$).

\noindent{\bf Remark:} As $k$ increases, logistic term in $G(.)$ forms a sharper transition around \emph{central point} \footnote{Value which makes logistic term equals to $\frac{1}{2}$.}. Therefore, for a sufficiently large $k$, $D$ would be negative if $\frac{m\alpha}{1+\rho_b} > m(1-p_i)$ 
\footnote{Mean of $\mathcal N(m(1-p_i),mp_i(1-p_i))$ $<$ central point of logistic coefficient of $\mathcal N(m(1-p_i),mp_i(1-p_i))$.} 
and $\frac{m\alpha}{1-\rho_b}<mp_i$ 
\footnote{Mean of $\mathcal N(mp_i,mp_i(1-p_i))$ $>$ central point of logistic coefficient of $\mathcal N(mp_i,mp_i(1-p_i))$.}. 
The aforementioned conditions are equivalent to $p_i > \max(\frac{\alpha}{1-\rho_b},1-\frac{\alpha}{1+\rho_b})$.  

\noindent{\bf Determining $\Delta$}: The proper choice of $i$ can be obtained by finding the least integer $i$ where $D$ is negative. By denoting the appropriate $i$ as $i^*$, then $\Delta$ (presented in step 1) can be found by solving $\frac{c}{2}+\Delta = i^*$.  

All tuples $(\rho_n,\rho_b)$ for which the safety is guaranteed can be obtained by checking if there exists an $i$ such that $D<0$ for those values of $\rho_n$ and $\rho_b$.

\begin{figure}[t]
\centering
\includegraphics[trim=55mm 19mm 50mm 16mm,clip,width=.7\textwidth]{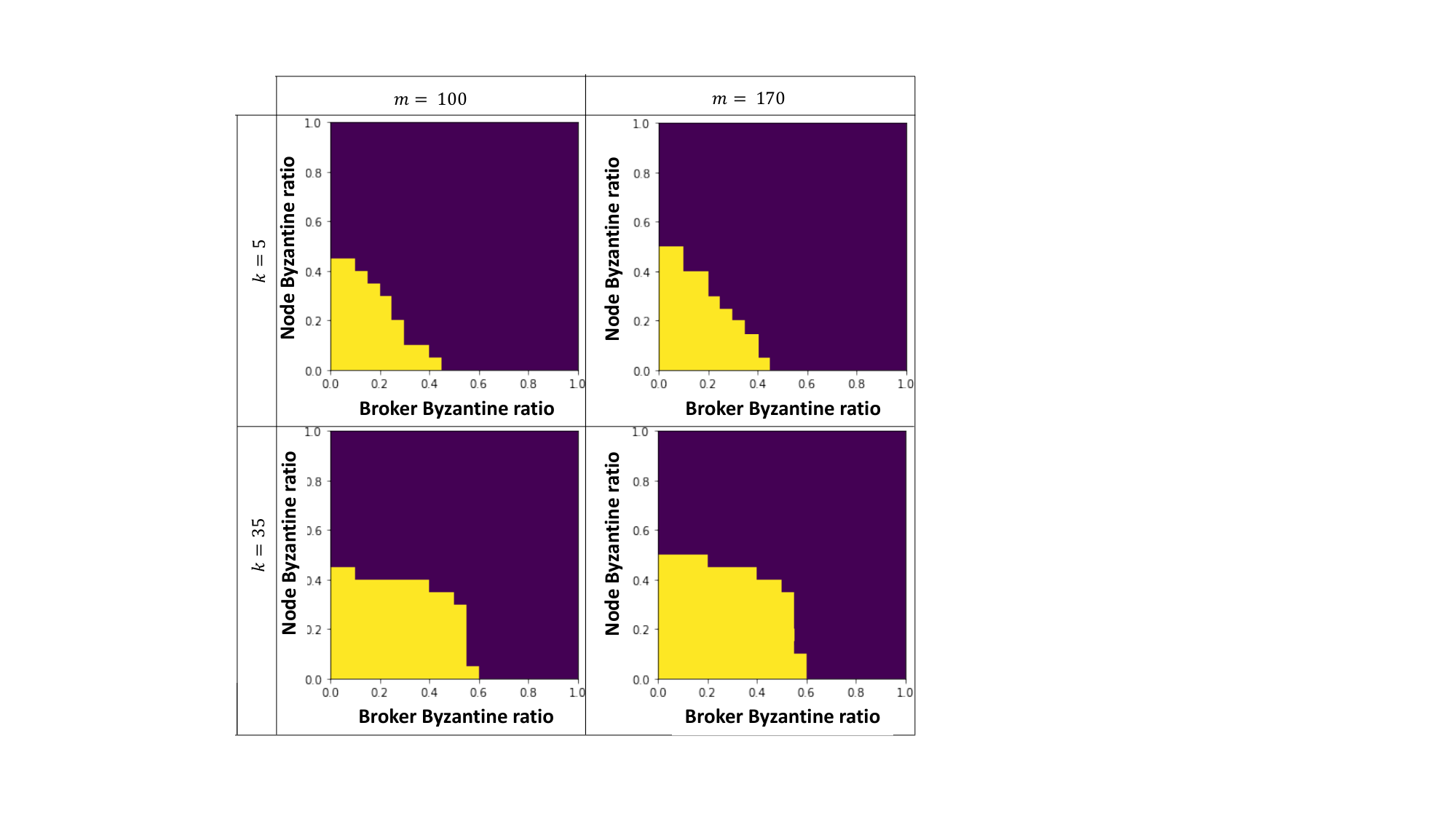}
\caption{An illustration of safety-guaranteed region (indicated with yellow color) of Blizzard protocol through simulation for different number of connections of each mobile node, i.e. $k$, and different number of brokers, i.e. $m$ while fixing total number of mobile nodes $n =2000$ and 1000 iterations. }
\label{safety_intro_fig}
\end{figure}

We further perform simulations to obtain all tuples $(\rho_n,\rho_b)$ such that the safety is assured for the case of having 2,000 mobile devices. Fig. \ref{safety_intro_fig} illustrates this guaranteed safety region, shown with yellow color, for different number of brokers and different number of connections. We consider 1,000 iterations with 0.05 resolution for the Byzantine ratio of nodes and brokers. One interesting observation is that Byzantine ratio of mobile nodes and brokers can respectively reach \%50 
(when $\rho_b$ is small) and $<$ 60\% of Byzantine brokers (when $\rho_n$ is small), with a tradeoff seen between these ratios. While the latter may seems surprising at first glance, it reflects the fact that the brokers in Blizzard are inherently much weaker in what they can do - they can only relay or suppress node messages.
\subsection{Liveness}
As with other DAG-based protocols such as IOTA~\cite{Iota} and Avalanche \cite{avalanche19}, liveness failure in Blizzard occurs when either a transaction has a invalid transaction as its parent or a transaction does not gain enough confidence value. The former scenario can be resolved by re-issuing the transaction with new valid parents, while the latter could be resolved by having a node send additional valid transactions as successors to increase the confidence value.


\section{Performance Analysis}
We first present our analysis on throughput per shard, then we focus on analyzing latency. 

\subsection{Throughput per Shard}\label{throughput_analysis}
\begin{figure*}[t]
\centering
\includegraphics[trim=40mm 110mm 30mm 10mm,clip,width=\textwidth]{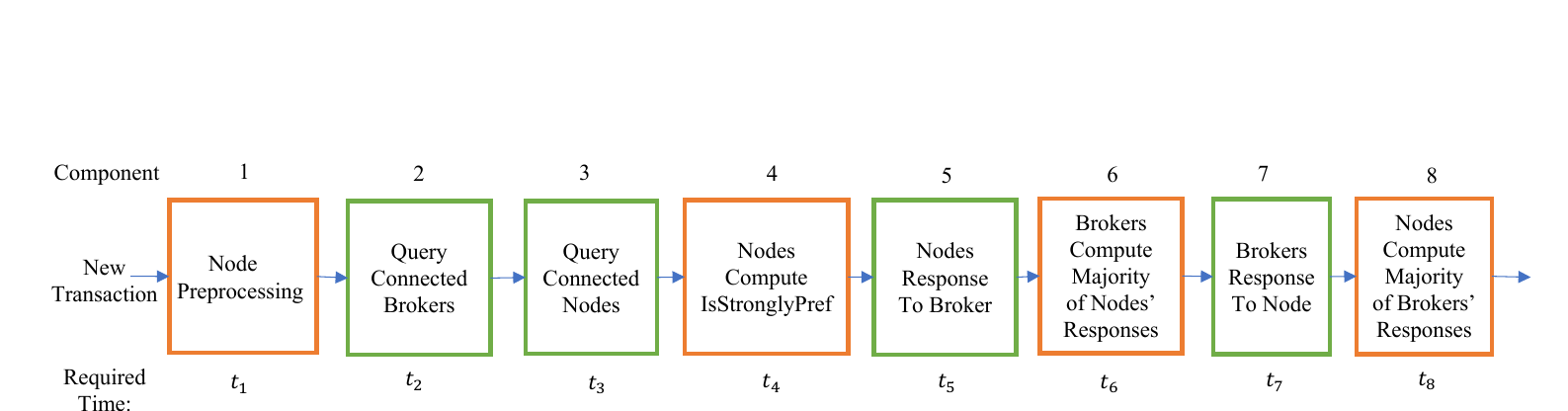}
\caption{The pipeline modeling of different components of Blizzard scheme to acquire throughput. We use orange and green colors for boxes as computing and communication components, respectively.} 
\label{fig_TPS}
\end{figure*}

We elaborate upon obtaining the throughput of Blizzard from a novel perspective, i.e. modeling it as a pipeline, as illustrated in Fig. \ref{fig_TPS}. By considering $t_i$ as the required time for performing the task of component $i$, and considering that the component with the smallest rate would dominate the result, throughput would be equal to $\min_{1\leq i\leq 8}\frac{1}{t_i}$.



By respectively considering the network bandwidth and transaction size as $BW$ bps and $300$ Bytes, the rate of communication components (green colored boxes), i.e. components 2-3, 5, and 7 would be around $\frac{BW}{2400}$ transactions per second (tps). Among the computing components (orange colored boxes), i.e. components 1,4,6,8, component 4 (checking if a transaction is strongly preferred) dominates the computing time due to the fact that it is more time-consuming compared to the other computing components. Using this analysis, we will quantify the throughput using experimental measurements in section~\ref{sec:implementation}. 



\subsection{ Latency}
Total latency is referred to time interval from a transaction is issued to it is finalized. Total latency is upper-bounded by the sum of three terms:
\small
\begin{equation}
\begin{aligned} 
\text{Latency} &\leq t_{\text{propagation}}  
+t_{\text{validation}} +t_{\text{confidence}}
\end{aligned}    
\end{equation}
\normalsize
These terms are, respectively, the 1) \emph{propagation time} $t_{\text{propagation}}$ - the time taken for a transaction to be disseminated to the whole network, 2) \emph{transaction validation time} $t_{\text{validation}}$,  and 3) the \emph{confidence-gathering time} $t_{\text{confidence}}$, which is the time required for a transaction to achieve a threshold level of confidence through successive transactions voting for it. We analyze each of the aforementioned terms as follows.

\subsubsection{Propagation Time}
Since the propagation time is linearly proportional to the number of communication rounds \footnote{A communication round is a communication transmission from a mobile node to a broker or vice versa.}, we aim to obtain the Least Number of Communication Rounds (LNCR) required for a transaction to propagate in the entire network by starting from the node which issued this transaction and ending by the last node which discovers this transaction. We are able to prove a strong result about LNCR in Blizzard. 

\begin{theorem}\nonumber
Let us assume that $\frac{(m-k)n}{(m-1)m}>1$, then LNCR of Blizzard equals 4 with high probability.
\end{theorem}

\begin{proof}
    We first prove that the distance between any two vertices which represent brokers on the corresponding bipartite graph is 2 with high probability. Then, the distance between any two vertices indicating mobile nodes would be at most 2 more than that, i.e. 4 with high probability. The probability that vertices $v_1$ and $v_2$, $\forall v_1\neq v_2$ and $v_1,v_2\in V$, have distance 2 can be obtained as follows:
    \small
    \begin{equation}\nonumber
    \begin{aligned}
        &P(dist(v_1,v_2)=2)=P(\text{existence of at least one node}\\&~~~ \text{$u\in U$ which is connected to both $v_1$ and $v_2$})  
    \\&\overset{(a)}{=}
    1-\Big(1-\frac{{{m-2}\choose {k-1}}}{{{m-1}\choose{k-1}}}\Big)^{\frac{n}{m}}
    =1-\Big(1-\frac{m-k}{m-1}\Big)^{\frac{n}{m}}
    \overset{(b)}{\approx}
    1-e^{-r}
    \end{aligned}    
    \end{equation}
    \normalsize
    where $(a)$ follows from considering average $\frac{n}{m}$ mobile nodes per broker and $(b)$ follows from $\lim_{n \to\infty}{(1-\frac{r}{n})^n}=e^{-r}$ and considering $r:= \frac{(m-k)n}{(m-1)m}$. It is easy to see $e^{-r}$ is small enough if $r>1$ (or equivalently the condition of the Theorem holds). This condition is realistic due to the fact that we expect $n>>m$.
\end{proof}

\subsubsection{Transaction Validation Time}
As mentioned earlier, transaction validation time refers to the time required for a node to check the validity of transactions. Therefore, we can use the pipeline scheme, explained in previous section on throughput and illustrated in Fig.~\ref{fig_TPS}, to model this time. Mathematically, transaction validation time can be expressed as $\sum_{i=1}^{8}{t_i}$.

\subsubsection{Confidence-gathering Time}
Let us assume $L$ represents the average number of transactions come later after a transaction appended to the DAG until it gets finalized\footnote{The value of $L$ would depend on the security parameters of the protocol as well as the DAG-attachment policy adopted by nodes; in the special case when all transactions are attached sequentially in a chain, it can be shown that $L$ would be equal to $\min{(\beta_1,\beta_2)}$.}. By defining $\zeta$ as the arrival rate of transactions, the average confidence-gathering time would be $\frac{L}{\zeta}$.     


Using the above analysis, we will quantify the latency using experimental measurements in section~\ref{sec:implementation}. 

\subsection{Average Message Complexity}

\emph{Average message complexity} refers to average number of messages required for a transaction to be queried by all nodes. The average message complexity of Blizzard can be obtained by noting that each broker is queried by one of its connected nodes and then collects and sends back the majority vote to all its connected nodes. This implies that Blizzard needs ${m}+2kn$ messages for querying.

\begin{figure}
\centering
\includegraphics[trim=105mm 62mm 102mm 60mm,clip,width=.55\textwidth]{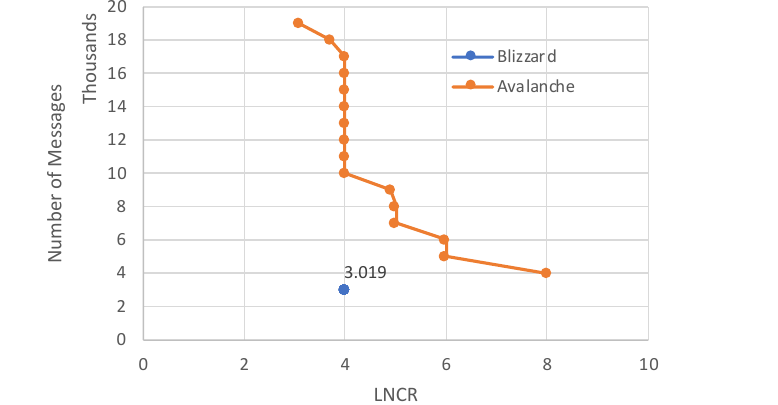}
\vspace{-.2in}
\caption{Numerical simulation-based comparison of total number of massages versus lowest number of communication rounds (LNCR) of Avalanche with Blizzard for the following setting: $n= 500$ and $k=3$, $m$ is varied from 4 to 19.}
\label{trafeoff_1fig}
\end{figure}

While we argue that Avalanche cannot be implemented on mobile device networks in a scalable manner as it requires direct peer to peer communication between any two random devices, we can still compare Blizzard and Avalanche in terms of their trade-off between message complexity and LNCR assuming Blizzard were to be implemented on the same network as Avalanche, as shown in Fig. \ref{trafeoff_1fig}. 
For having a fair comparison (i.e. equal  number  of  nodes  being  queried  on each  transaction), $q$ (number of sampled nodes in Avalanche) should be $\frac{nk}{m}$.
As $q$ increases in Avalanche protocol, LNCR decreases while total number of required messages significantly increases. 
However, Blizzard protocol obtains the best of both worlds, i.e. having a lower LNCR and lower total number of required messages \footnote{Note that number of required messages in Blizzard could be reduced further by decreasing $m$ but this would result in lower security.}.

\section{Implementation and Experimental Measurements \label{sec:implementation}}
We implemented Blizzard in C++ and also a version of Avalanche for comparison purposes. We have made our source code for the implementation of Blizzard as well as its comparison with Avalanche publicly available at \url{https://github.com/ANRGUSC/Blizzard}.

Using our implementation, we ran computations of Blizzard on a compute machine with configuration of 2.3 GHz Intel core i5 (which its frequency is less than the the frequency of state-of-the-art Mobile CPUs \cite{mobile_power_frequency}), so that all computations are emulated in real time. For the communication between nodes and brokers, we simulated it with one-way network latency drawn from uniform distribution with 100ms mean and standard deviation 25ms. We next present the experimental measurements of throughput and latency, as well as an estimation of battery energy consumption. 

\subsection{Throughput Evaluation}
As we explained in \ref{throughput_analysis}, since component 4 (checking if a transaction is strongly preferred) in Fig. \ref{fig_TPS} dominates the computational time, we implemented component 4 in C++ and observed empirically that $t_4 \approx 100\mu s$ for the setting where there are 400 transactions known by mobile nodes. Since the computing power of top 10 iOS mobile devices exceeds the computing power of the device used in the implementation we performed\footnote{Based on https://www.geekbench.com/ }, the throughput of Blizzard is as presented in the following table: 

\begin{center}
\begin{tabular}{cccc} 
\toprule
{\shortstack{Network\\Bandwidth}} & $100$ Mbps & $10$ Mbps & $1$ Mbps\\
\midrule
\shortstack{Throughput\\on PCs} & 10,000 TPS & 4,166 TPS & 416 TPS\\
\bottomrule
\label{TPS_TABLE_2}
\end{tabular}
\end{center}

\subsection{Latency Evaluation}
Assuming security parameters $\beta_1=11$, $\beta_2=150$, a chain topology for the DAG and a transaction arrival rate higher than 100 tps, the propagation and validation times are going to be dominant compared to the confidence time, and in turn they will each be dominated by four communication steps; this implies a total latency on the order of $<1$s ($\sim 0.65$s).

\begin{figure*}[tbh]
\centering
\includegraphics[scale=0.57]{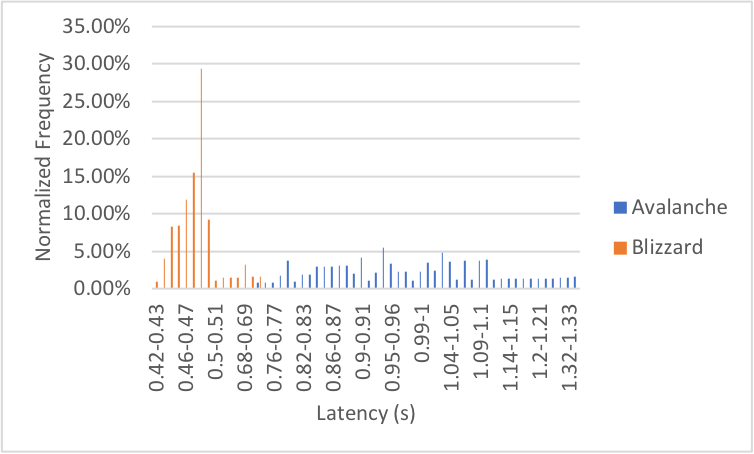}
\includegraphics[scale=0.57]{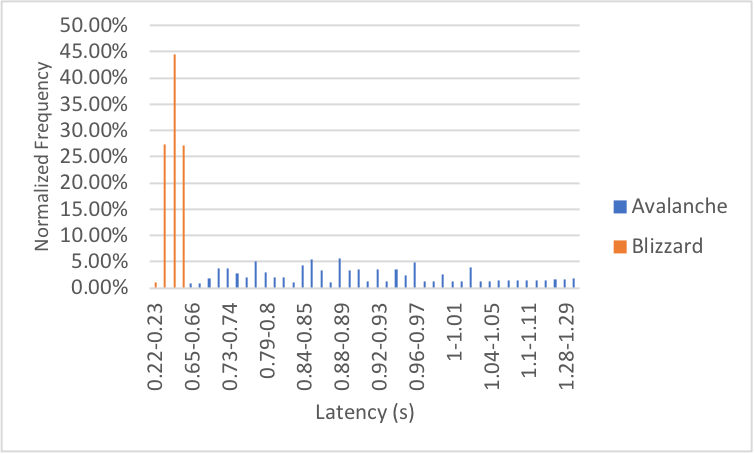}
\caption{ The left and right plots respectively represent the histogram of transactions latency of our proposed scheme (Blizzard) compared with Avalanche \cite{avalanche19} for the case of 100 transactions, security parameters $\beta_1=11$ and $\beta_2=150$. The left histogram corresponds to the settings of 100 nodes, 8 brokers, and 3 connections per node; while the right figure corresponds to the case of 200 nodes, 11 broker, and 6 connections per node.}
\label{fig:latency_histogram}
\end{figure*}

Fig. \ref{fig:latency_histogram} shows the histograms of transactions latency of our proposed scheme and Avalanche \cite{avalanche19} in two different settings. As it can be observed, Blizzard significantly reduces latency by $\sim 50\%$. Further, one can see Avalanche has a wide range of transactions latency while Blizzard has a dense one.       

\subsection{Battery Energy Consumption}

We give here a rough estimation of the energy consumed per transaction validated by a mobile device. As the amount of data communicated between the mobile nodes and online brokers for each transaction is relatively small, and further, no computationally expensive Sybil control mechanism like Proof of Work is required, the primary source of energy consumption for our protocol is computation. As discussed above, the dominant source of computation when validating a transaction (see Figure \ref{fig_TPS}) is component 4 (Computing IsStronglyPref.). Based on our experiments we estimate this takes about 100 $\mu$s. Assuming a mobile CPU power consumption of 1.5 Watt \cite{mobile_power_frequency} and conservatively assuming full CPU utilization, this would translate to about $1.5 \times 10^{-4}$ Joules per transaction. While we are not aware of benchmark numbers for other protocols we could compare this with since other protocols are typically not built with energy efficiency of validators in mind, we believe this imposes a relatively low, manageable load on a mobile device, particularly as the device owner is free to determine what number of transactions it participates in over a given period of time.

\section{Discussion}
{Here we briefly elaborate on topics which merit further attention. These topics are out of scope for the present paper, but we are actively pursuing these directions.

\subsection{Mobile-Device oriented Sybil Control}
What has been presented in this paper thus far is a consensus mechanism. It implicitly assumes that there is already a Sybil-control mechanism in place, such as one based on Proof of Work or Proof of Stake. To implement Blizzard in a network with millions of mobile devices, it may be helpful to create a Sybil control mechanism such that only users with valid mobile devices can participate in the consensus with significant cost associated with creating or operating multiple, potentially fake identities. 
A potential design for such a system could leverage the existence of globally unique mobile ID's such as IMEI numbers, while still maintaining an overall architecture that is sufficiently open and decentralized.  Another alternative is to utilize decentralized IDs on the mobile nodes with a permissioned setup. Yet another approach may be to use location information or wireless signal strength as the basis for a Sybil control mechanism~\cite{jiang2018senate}-\cite{king2018introduction}. Another approach could be to use Proof of Social Contacts~\cite{proofofsocialContacts}, which leverages encounter information between mobile nodes to detect and blacklist Sybil nodes.

\subsection{Improving Scalability}
One of the crucial bottleneck of a DAG-based protocol is that all nodes need to store the entire DAG as well as investigating whether a transaction is strongly preferred by going through entire DAG. As a result, the system faces storage and computation bottlenecks. These challenges may be exacerbated when involving relatively more resource-limited mobile nodes in the consensus mechanism, as we have proposed in Blizzard. To address the computation and storage challenge, we consider three approaches, namely \emph{sharding}~\cite{sharding}, \emph{pruning DAG}, and \emph{off-loading verification}, to improve the scalability.

\noindent{\bf Sharding}: This technique creates multiple pools of mobile nodes, where each pool focuses on storing and verifying transactions belonging to a corresponding subset of all accounts (per the well-known Blockchain Trilemma, this solution trades off security for scalability while maintaining decentralization). 

\noindent{\bf Pruning DAG}: By defining \emph{check point transaction} as the one which is finalized, it is clear that all ancestor transactions of a check point transaction is also finalized. In the account model (the state-based approach used in Ethereum)  as long as we reach a check point transaction on the DAG, we do not need to store all its ancestor transactions. Therefore, each mobile node can save significant amount of memory by storing only the pruned DAG. By including such a concept of check points, we could fulfill the criterion of having lightweight nodes. However, as discussed in~\cite{pruning} there are a number of other practical considerations that should be kept in mind, such as ensuring historical information is retained on some full nodes and accessible in a decentralized manner for security purposes. In our architecture, the online brokers could potentially serve the role of full nodes that store the entire history, while the mobile nodes only store information past the last checkpoint.

\noindent{\bf Off-loading verification}:
Inspired by \cite{offloading_vitalik}, this approach would allow the computation associated with verification (investigating whether a transaction is strongly-preferred or not) to be offloaded to more powerful servers that provide zero knowledge proofs which can be verified in a more lightweight manner by the mobile devices. More research is needed to flesh out and realize such an approach. } 

\subsection{Safety under a Partially Synchronous Model}
As shown in~\cite{permissionless_sync} in order to prove safety under a partially synchronous model, either the protocol would have to explicitly take into account the total number of participating nodes and their votes in determining when to finalize a transaction (such as done in protocols like Tendermint \cite{Tendermint} and Hotstuff \cite{Hotstuff}). Alternatively, it would be interesting to explore the development of a decentralized BFT approach to empirically quantifying (possibly in a time-varying manner) an upper bound on the network latency at all times. Given such a mechanism, the protocol parameters could be suitably adapted to ensure correct and efficient operation despite a time-varying network latency, i.e. in a partially synchronous network.

\section{Conclusion}
In this work, we have presented Blizzard, mobile-device oriented BFT consensus-based distributed ledger protocol. Blizzard incorporates a novel two-tier broker-based architecture and a decentralized random matching mechanism. We have mathematically analyzed and presented the safety guarantee for Blizzard. Interestingly this guarantee is in the form of a two-dimensional region -- for sufficiently large networks, our numerical computations show that the protocol is capable of supporting $<$ 50\% of Byzantine nodes (when the number of Byzantine brokers is small) and $<$ 60\% of Byzantine brokers (when the number of Byzantine nodes is small), with a tradeoff seen between these ratios. 

We have also discussed how Blizzard satisfies liveness. Moreover, we have also analyzed and evaluated the performance of Blizzard in terms of throughput, latency, and message complexity.  
We also showed that Blizzard has superior performance in terms of significant low message complexity and short propagation latency compared to Avalanche \cite{avalanche19}, which in any case would be challenging to implement in a mobile-first scenario as we consider here as it is challenging to allow mobile devices to directly randomly query any other mobile nodes in a large network without going through any online servers.   

We have shown that Blizzard can provide an acceptable level of throughput per shard. To improve throughput performance further, it is important to develop or adopt additional scaling mechanisms that have been proposed in other projects such as account-based sharding and second-layer solutions such as state channels.  Memory limitations of mobile devices can be addressed by a suitable combination of check-point-based DAG pruning and off-loaded verification~\cite{offloading_vitalik} solutions. As a key future direction of this work, we aim to implement Blizzard scheme on real mobile devices and empirically measure throughput and latency.

\appendices
\section{High Probability Connections}\label{appendix_thm}
By defining random variable $X$ as number of Byzantine brokers picked by selecting $\ell$ brokers from all brokers, then $P(X=x)=\frac{ {{m_b}\choose{x}} {{m_c}\choose{\ell-x}} }{ {{m_b+m_c}\choose{\ell}}}$. Based on Hoeffding inequality, we have $P(X-{\mathbb E}[X]>\theta)\leq e^{-\frac{2\theta^2}{{\ell}^2} }$ where ${\mathbb E}[X]=\frac{m_b}{m}\ell=\rho_b \ell$. 
 
\section{Overview on DAG Part of Avalanche:}\label{overview_dag}

A mobile node designates several parents for a new transaction once upon issuing a new transaction and forms edges on the DAG. The main difficulty of keeping the DAG is to select one transaction among conflicting ones. Double-spending is one of the examples of conflicting transactions. 
Once upon a transaction is queried, all ancestor transactions of this one is implicitly included in the query. A node affirmatively responds to the query if all of the ancestors are currently the \emph{preferred} choice of transaction in their corresponding conflict sets. In the case of having $\geq \alpha k$  (where $\frac{1}{2}<\alpha<1$) positive responds for transaction $T$, then this transaction receives voucher $v_{u,T} = 1$ and be appended to the DAG; otherwise $v_{u,T} = 0$.    

Every node stores the entire transactions it has known in its DAG. Each DAG consists of mutually exclusive conflict sets ${\mathcal P}_T$ for $T\in {\mathcal T}_u$ where ${\mathcal T}_u$ represents the subset of known transaction by node $u$. Each conflict set ${\mathcal P}_T$ has three components, namely the preferred transaction ${\mathcal P}_T.\text{pref}$, last seen transaction ${\mathcal P}_T.\text{last}$, and counter ${\mathcal P}_T.\text{counter}$. Moreover, every node $u$ computes the \emph{confidence} value of transaction $T$ by the following formula:
\begin{equation}
    d_{u}(T) \triangleq \sum_{T': T' \in {\mathcal T}_u, T'\rightsquigarrow T}{v_{u,T'}}
\end{equation}
where $T' \rightsquigarrow  T$ indicates a path from $T'$ to $T$. Furthermore, DAGs created by different nodes are assured to be consistent, meaning that relation $T \rightarrow  T'$ exists for the DAG of all nodes if $T \rightarrow  T'$; and there is no node with relation $T \rightarrow  T'$ if $T \not\rightarrow  T'$. 
Please see ~\cite{avalanche19} for transaction finalization.


\section*{Acknowledgment}
This work was supported in part through a gift to the USC Center for Cyberphysical Systems and the Internet of Things from SovereignWallet Network.

\ifCLASSOPTIONcaptionsoff
  \newpage
\fi



%




\bibliographystyle{IEEEtran}
\bibliography{references}

\end{document}